\documentclass[pra,aps,amssymb,twocolumn,noshowpacs]{revtex4-1}
\usepackage{color}
\usepackage{graphicx}
\usepackage{hyperref}
\usepackage{amsmath}
\usepackage{latexsym}
\usepackage{amssymb}
\usepackage{mathrsfs}
\usepackage{mathtools}
\usepackage{layout}
\usepackage{verbatim}
\usepackage{multirow}
\usepackage{bm}
\usepackage{amsfonts,epsfig}

\begin{document}

\title{Suppress motional dephasing of ground-Rydberg transition for high-fidelity quantum control with neutral atoms}

\date{\today}
\author{Xiao-Feng Shi}
\affiliation{School of Physics and Optoelectronic Engineering, Xidian University, Xi'an 710071, China}

\begin{abstract}
	The performance of many control tasks with Rydberg atoms can be improved via suppression of the motion-induced dephasing between ground and Rydberg states of neutral atoms. The dephasing often occurs during the {\it gap} time when the atom is shelved in a Rydberg state before its deexcitation. By using laser fields to induce specific extra phase change to the Rydberg state during the gap time, it is possible to faithfully transfer the Rydberg state back to the ground state after the gap. Although the Rydberg state transitions back and forth between different eigenstates during the gap time, it preserves the blockade interaction between the atom of interest and a nearby Rydberg excitation. This simple method of suppressing the motional dephasing of a flying Rydberg atom can be used in a broad range of quantum control over neutral atoms.

\end{abstract}
\maketitle

\section{introduction}\label{sec01}
The study of Rydberg interaction has drawn intense interest in recent years due to its applicability in quantum information~\cite{Saffman2010,Saffman2016,Weiss2017} and quantum optics~\cite{Dudin2012,Peyronel2012,Firstenberg2013,Li2013,Gorniaczyk2014,Baur2014,Tiarks2014,Busche2017,Lampen2018,Adams2019}. Unfortunately, Rydberg excitation is usually achieved via using lasers, but nothing is static in this world so that even if an atom is cooled to the order of $10~\mu K$, its root-mean-square~(r.m.s.) velocity $v_{\text{rms}}$ along the laser propagation direction is still too large as seen by the atom and the photon interacting with the atom. This motion induces significant dephasing characterized by $v_{\text{rms}}$ and the effective wavevector, $k$, of the Rabi frequency of the Rydberg lasers. The larger $k$ and $v_{\text{rms}}$, the more severe the dephasing. To minimize the dephasing, the excitation of an $s~(d)-$orbital Rydberg state is achieved via a two-photon transition with the lower and upper laser fields propagating in opposite directions. However, it still leads to a $k$ of several times $10^6m^{-1}$ if the element used is $^{87}$Rb or $^{133}$Cs, as frequently adopted in experiments~\cite{Wilk2010,Isenhower2010,Zhang2010,Maller2015,Zeng2017,Levine2018,Picken2018,Levine2019,Graham2019}. This leads to a $1/e$ dephasing time $\propto1/(kv_{\text{rms}})$ ~\cite{Wilk2010,Jenkins2012,Saffman2011,Graham2019} of only several microseconds in typical experimental setups~\cite{Wilk2010,Isenhower2010,Saffman2011,Dudin2012,Graham2019}. Although a method is given in Ref.~\onlinecite{Ryabtsev2011} for a Doppler-free three-photon transition between ground and Rydberg states of a rubidium atom, to the best of our knowledge, however, there is no theory for suppressing the dephasing in the usually adopted two-photon transition~\cite{Wilk2010,Isenhower2010,Zhang2010,Maller2015,Zeng2017,Levine2018,Picken2018,Levine2019,Graham2019}. 

The difficulty of eliminating the motional dephasing lies in the implementation of the same quantum control (usually by optical pumping) for all atomic velocity, $v$, in spite of the finite distribution of $v$ along the light propagation. The effect of this velocity distribution on the quantum control is analogous to inserting a linear process 
in a random precession which leads to fast dephasing. To exacerbate the problem, a gap is usually inserted between the excitation and deexcitation of the Rydberg state~\cite{Isenhower2010,Graham2019,Zhang2010,Maller2015,Zeng2017,Picken2018,Tiarks2014,Baur2014}, resulting in rapid dephasing. To eliminate this detrimental dephasing, a useful protocol should tailor the effect to each given $v$, but this seems exceedingly challenging since the same protocol should be implemented for all possible $v$.

In this work, we provide a general theory for suppressing the motional dephasing in a two-photon ground-Rydberg transition. Our theory is applicable whenever a gap time exists or can be inserted between the excitation and deexcitation of the Rydberg states, and is based on a basic property of the Doppler dephasing in the ground-Rydberg transition: it originates from a {\it regular} phase shift of the atomic state, where the phase change is linear 
in $v$ and the time of optical pumping. With an appropriate ratio between the amplitude and wavevector of the Rabi frequency of the laser fields, the atomic state not only can return from Rydberg to ground state, but also undo the drift-induced phase twist. This simple method for suppressing the motional dephasing can be used for many quantum control tasks with Rydberg atoms.

\section{A $\pi-2\mathcal{N}\pi-\pi$ protocol}
\begin{figure}
\includegraphics[width=3.2in]
{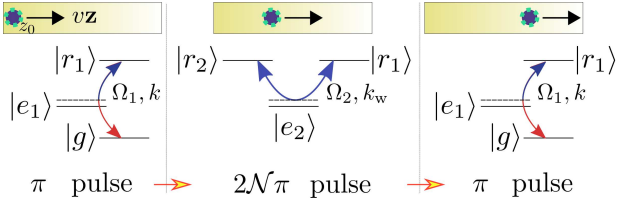}
 \caption{A $\pi-2\mathcal{N}\pi-\pi$ protocol for removing the motional dephasing of a Rydberg atom in free flight. Starting from the ground state $|\psi(0)=|g\rangle$, a $\pi$ pulse for the transition $|g\rangle\rightarrow|r_1\rangle$ takes the atom to $|r_1\rangle$, followed by a $2\mathcal{N}\pi$ pulse for the transition $|r_2\rangle\leftrightarrow|r_1\rangle$. A final $\pi$ pulse restores the atom to the ground state. The motional dephasing is suppressed if $k_{\text{w}}/k =2+\Omega_{\text{2}}/(\mathcal{N}\Omega_{\text{1}})$ and $v\ll$max$\{\Omega_1/k, \Omega_2/k_{\text{w}}\}$. $|e_1\rangle$ and $|e_2\rangle$ are low-lying intermediate states for the two-photon transitions $|g\rangle\leftrightarrow|r_1\rangle$ and $|r_2\rangle\leftrightarrow|r_1\rangle$, respectively. The solid lines denote atomic eigenstates, while the dashed lines indicate that the $|e_{1(2)}\rangle$ states are largely detuned so that they are barely populated. \label{figure01} }
\end{figure}

We describe our theory for a three-level system with a ground state $|g\rangle$ and two Rydberg eigenstates $|r_{1(2)}\rangle$ in a heavy alkali-metal element like $^{87}$Rb and $^{133}$Cs. Here, $|r_{1}\rangle$ and $|r_{2}\rangle$ denote two different $s~(d)-$orbital Rydberg states with principal quantum number of, for example, 100. Our method is implemented using three pulses: a $\pi$ pulse, a $2\mathcal{N}\pi$ pulse, and then a $\pi$ pulse, where $\mathcal{N}$ is an integer; the scheme is shown in Fig.~\ref{figure01}. The theory can be understood by observing the phase change of the atomic state of an atom with a drift velocity of, e.g., $v\mathbf{z}$; we assume that the drift along $\mathbf{x}$ and $\mathbf{y}$ does not change the Rabi frequency during the timescale of interest~\cite{Shi2018Accuv1} so that only the atomic coordinate along $\mathbf{z}$ is relevant to the problem. If the atom is initially at $z_0$ and in the internal state $|\psi(0)=|g\rangle$, a $\pi$ pulse with the Hamiltonian
\begin{eqnarray}
 \hat{H}_{\text{1}} &=& \Omega_{\text{1}} e^{ik(z_0+vt)} |r_1\rangle\langle g|/2+\text{H.c.}\label{equation01}
\end{eqnarray}
will take the atom to the Rydberg state $|\psi(t)\rangle=e^{i\varphi_1}|r_1\rangle$ at $t=t_{\pi}\equiv\pi/\Omega_1$, where $\Omega_1$ is the magnitude of the Rabi frequency~(similar for $\Omega_2$ below). It is easy to show that $\varphi_1=kz_0-\pi/2$ if $v=0$. For a nonzero $v$ bounded by $|v|\ll\Omega_1/k$, we find that $\varphi_1$ is given by 
\begin{eqnarray}
\varphi_1 &=&kz_0+kvt_{\pi}/2-\pi/2. \label{varphi01}
\end{eqnarray} 
If we cease the optical pumping of Eq.~(\ref{equation01}) at $t=t_{\pi}$ and switch on the following transition 
\begin{eqnarray}
 \hat{H}_{\text{2}} &=& \Omega_{\text{2}} e^{ik_{\text{w}}(z_0+vt)} |r_1\rangle\langle r_2|/2+\text{H.c.}.\label{equation02}
\end{eqnarray}
for a duration of $t_{\text{w}}=2\mathcal{N}\pi/\Omega_{\text{2}}$, the state evolves to $|\psi(t)\rangle=e^{i\varphi_2}|r_1\rangle$ at $t=t_{\pi}+t_{\text{w}}$, where
\begin{eqnarray}
\varphi_2 &=& \varphi_1+k_{\text{w}}vt_{\text{w}}/2-\mathcal{N}\pi,
\end{eqnarray} 
where $\mathcal{N}$ is a positive integer. One can show that by switching off Eq.~(\ref{equation02}) and switching on the pumping of Eq.~(\ref{equation01}) for another $\pi$ pulse, the state becomes $|\psi(t)\rangle=e^{i\varphi_3}|g\rangle$ at $t=2t_{\pi}+t_{\text{w}}$, where
\begin{eqnarray}
\varphi_3 &=& \varphi_2-kz_0-kv (t_{\text{w}}+3t_{\pi}/2 )-\pi/2.\label{varphi3}
\end{eqnarray} 
The above equation means that as long as there is a physical solution to
\begin{eqnarray}
  k_{\text{w}}/k &=&  2+\Omega_{\text{2}}/(\mathcal{N}\Omega_{\text{1}}) ,\label{condition01}
\end{eqnarray} 
$\varphi_3$ becomes exactly $\pi$ or $2\pi$ and thus one has removed the effect of the random velocity in the optical pumping process, i.e., the Doppler dephasing. In the derivation above, we have assumed both $\Omega_{\text{1}}$ and $\Omega_{\text{2}}$ to be positive, thus any $k_{\text{w}}/k$ that is larger than $2$ can 
satisfy Eq.~(\ref{condition01}). Because the transitions $|g\rangle\leftrightarrow|r_1\rangle$ and $|r_2\rangle\leftrightarrow|r_1\rangle$ are realized by two-photon excitation, their intermediate states, $|e_1\rangle$ and $|e_2\rangle$, determine $k_{\text{w}}/k$. Table~\ref{table1} lists eight choices of $|e_{1(2)}\rangle$ for $^{87}$Rb or $^{133}$Cs when both $|g\rangle\leftrightarrow|r_1\rangle$ and $|r_2\rangle\leftrightarrow|r_1\rangle$ are implemented by counterpropagating fields. Besides, several other choices like using $6P_{1(3)/2}$~(or $7P_{1(3)/2}$ and $8P_{1(3)/2}$) as $|e_{1}\rangle$ for $^{87}$Rb~(or $^{133}$Cs) can also be used but not listed in Table~\ref{table1}.

\begin{table}
  \centering
  \begin{tabular}{|c|c|c|c|c|c|c|c|c|c|}
    \hline
    \multirow{3}{*}{$^{87}$Rb}&   $|e_1\rangle$ &   \multicolumn{4}{|c|}{$5P_{\frac{1}{2}}$} &   \multicolumn{4}{|c|}{$5P_{\frac{3}{2}}$}\\
    \cline{2-10}
    &  $|e_2\rangle$ & $5P_{\frac{1}{2}}$& $5P_{\frac{3}{2}}$& $6P_{\frac{1}{2}}$& $6P_{\frac{3}{2}}$& $5P_{\frac{1}{2}}$& $5P_{\frac{3}{2}}$& $6P_{\frac{1}{2}}$& $6P_{\frac{3}{2}}$\\
    \cline{2-10}
    & $k_{\text{w}}/k$ & 4.95&4.90 &2.34&2.32 &5.25 &5.19 &2.48 &2.46 \\\hline
    \multirow{3}{*}{$^{133}$Cs}&   $|e_1\rangle$ &   \multicolumn{4}{|c|}{$6P_{\frac{1}{2}}$} &   \multicolumn{4}{|c|}{$6P_{\frac{3}{2}}$}\\
    \cline{2-10}
    &  $|e_2\rangle$ & $6P_{\frac{1}{2}}$& $6P_{\frac{3}{2}}$& $7P_{\frac{1}{2}}$& $7P_{\frac{3}{2}}$& $6P_{\frac{1}{2}}$& $6P_{\frac{3}{2}}$& $7P_{\frac{1}{2}}$& $7P_{\frac{3}{2}}$\\
    \cline{2-10}
    & $k_{\text{w}}/k$ &4.47 & 4.35&2.13 & 2.09& 5.10& 4.96& 2.43& 2.38\\\hline    
  \end{tabular}
  \caption{ Choices of the intermediate state $|e_{1(2)}\rangle$~(see Fig.~\ref{figure01}) for achieving $k_{\text{w}}/k>2$ so that Eq.~(\ref{condition01}) can be fulfilled. Here $k=2\pi|\lambda_{\text{up}}^{-1}-\lambda_{\text{lower}}^{-1}|$, where $\lambda_{\text{up}}$ and $\lambda_{\text{lower}}$ are the wavelengths of the upper and lower transitions~[see Fig.~\ref{figure01}(a)], respectively, and $k_{\text{w}}=4\pi/\lambda_{\text{w}}$, where $\lambda_{\text{w}}$ is the wavelength of the laser for the transition $|e_2\rangle\leftrightarrow|r_{1(2)}\rangle$. Data are obtained according to Refs.~\onlinecite{Sansonetti2006,Sansonetti2009} and Rydberg states with principal quantum number around 100~(lower states can also be used).  \label{table1}  }
  \end{table}


\begin{figure}
\includegraphics[width=2.8in]
{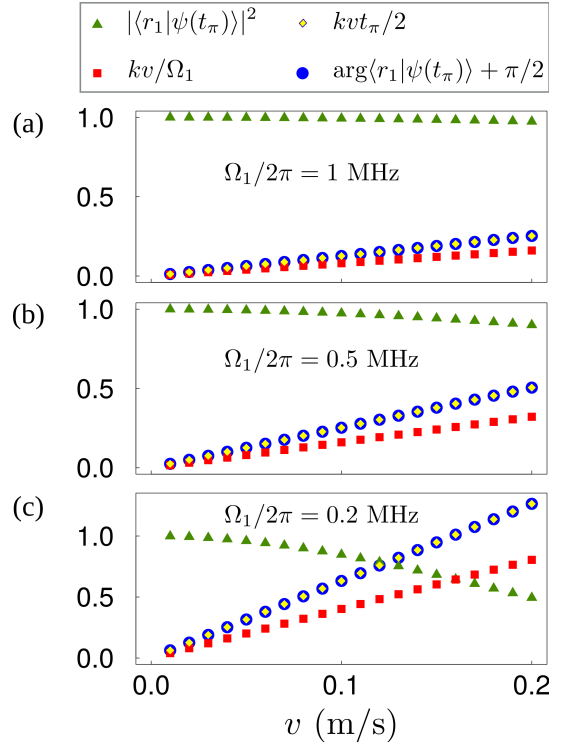}
 \caption{Population and phase of the Rydberg state $|r_1\rangle$ after using Hamiltonian~(\ref{equation01}) upon the initial state $|\psi(0)=|g\rangle$ with a $\pi$ pulse. The triangles show the population $|\langle r_1|\psi(t_\pi)\rangle|^2$, the filled circles show arg$\langle r_1|\psi(t_\pi)\rangle+\pi/2$, which coincides with $kvt_{\pi}/2$ with relative errors below $10^{-10}$ according to the precision of computers used in simulation. The ratio between $kv$ and $\Omega_1$ is shown by the square symbols. $\Omega_1/2\pi$ is $1,~0.5$, and $0.2$~MHz in (a), (b), and (c), respectively. The intermediate state $|e_1\rangle$ is the $5P_{3/2}$ state of $^{87}$Rb and $z_0=0$ is used.   \label{figure02} }
\end{figure}

Equation~(\ref{varphi01}) is the central equation in our theory. Its validity hinges on $v\ll\Omega_1/k$. For a numerical validation, we consider a rubidium-87 atom where the $5P_{3/2}$ state is used as the low-lying state for the two-photon transition between the ground state and a Rydberg state with a principal quantum number around 100. The wavelengths of the lower and upper laser fields are $780$ and $479$~nm, respectively, giving $k=5.06\times10^6/m$. The condition $v\ll\Omega_1/k$ means that $\Omega_1/2\pi$ should be at least $1$~MHz for atoms cooled to $10~\mu$K if high-fidelity quantum control is desired. The performance of the scheme is shown in Figs.~\ref{figure02}(a),~\ref{figure02}(b), and \ref{figure02}(c) with Rabi frequencies $1,~0.5$, and $0.2$~MHz, respectively. In Fig.~\ref{figure02}, $z_0=0$ is used because a nonzero $z_0$ does not induce any nontrivial effect. If $v$ corresponds to $v_{\text{rms}}\equiv\sqrt{k_BT/m}$, where $k_B$ is the Boltzmann constant, $T$ the atomic temperature, and $m$ the atomic mass, the corresponding $T$ changes from $1$ to $400~\mu$K when $v$ changes from $0.01$ to $0.2$~m/s in Fig.~\ref{figure02}. The result in Fig.~\ref{figure02}(a) agrees well with Eq.~(\ref{varphi01}), showing that a Rabi frequency of $1$~MHz is good enough.

One common feature in Fig.~\ref{figure02} is that, if we ignore the population error in $|r_1\rangle$, the phase of  $\langle r_1|\psi(t_\pi)\rangle$ is exactly given by Eq.~(\ref{varphi01}). However, our task is to populate the Rydberg state and then depopulate it, thus the population error should be avoided. In the numerical simulation, we find that the population $|\langle r_1|\psi(t_\pi)\rangle|^2$ is greater than $0.99$ when $kv/\Omega_1\leq0.1$, which corresponds to a threshold drifting velocity of $v\approx0.12$~m/s if $\Omega/2\pi=1$~MHz. This shows that high-fidelity control over a flying Rydberg atom is achievable even if the atom is relatively ``hot'' with a temperature of around $0.1$~mK.

\begin{figure}
\includegraphics[width=3.2in]
{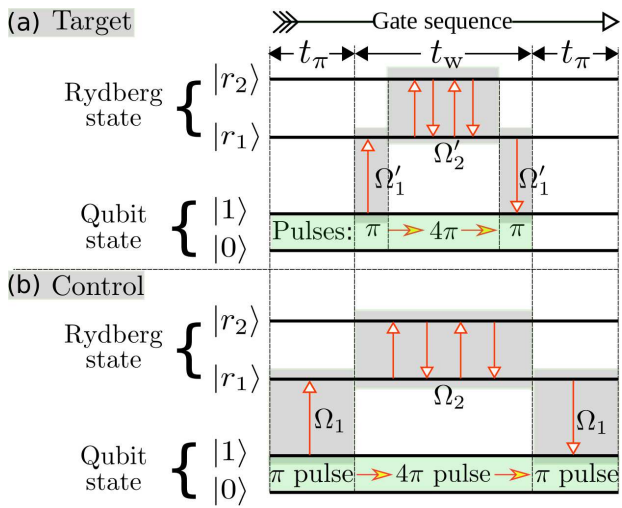}
 \caption{ Scheme of a Rydberg blockade gate by the $\pi-2\mathcal{N}\pi-\pi$ pulse sequence, where $\mathcal{N}$ is 2 for both qubits. The horizontal lines denote atomic eigenstates, and vertical arrows denote rotations between atomic states.  \label{figure03} }
\end{figure}

\section{application in Rydberg blockade gate}
To show that the above theory can be used in high-fidelity quantum control, we take the example of the Rydberg blockade gate~\cite{PhysRevLett.85.2208}, which maps the input states according to $\{|00\rangle,~|01\rangle,~|10\rangle,~|11\rangle\}\rightarrow \{|00\rangle,~-|01\rangle,~-|10\rangle,~-|11\rangle\}$, where $|\alpha\beta\rangle\equiv |\alpha\rangle_{\text{control}}\otimes|\beta\rangle_{\text{target}}$ with $\alpha$ and $\beta$ denoting the qubit states 0 or 1. The original protocol consists of three pulses: (i) the control qubit, if initialized in $|1\rangle$, is pumped to the Rydberg state $|r_1\rangle$ by a $\pi$ pulse; (ii) the target qubit, if initialized in $|1\rangle$, experiences a $2\pi$ pulse for the transition $|1\rangle\leftrightarrow|r_1\rangle$; (iii) a final $\pi$ pulse is used in the control qubit for $|r_1\rangle\rightarrow|1\rangle$. The free flight of the qubits during the sequence imposes a severe Doppler dephasing~\cite{Wilk2010,Saffman2011,Graham2019}. 

To suppress the dephasing, our $\pi-2\mathcal{N}\pi-\pi$ pulse sequence can be used in both the control and target. To induce entanglement, $\mathcal{N}$ can be either odd or even for the control qubit, but can only be even in the target. The scheme is shown in Fig.~\ref{figure03}, where $\mathcal{N}$ is, 
for illustration, $2$ for both qubits. The protocol in Fig.~\ref{figure03} maps the states similarly to the original gate. Before a numerical verification of the suppression of the Doppler dephasing, we first study the gate error from Rydberg-state decay below.

For brevity, we start from the analysis of the control qubit according to the pulse sequence in Fig.~\ref{figure03}. The gate duration is given by
\begin{eqnarray}
  t_{\text{gate}}   &=&\frac{2\pi}{\Omega_1}\frac{k_{\text{w}}-k}{k_{\text{w}}-2k}\label{gatetime}
\end{eqnarray}
according to Eq.~(\ref{condition01}). To minimize $t_{\text{gate}}$  for a given $\Omega_1$, it is better to use largest possible $k_{\text{w}}/k$ . Table~\ref{table1} shows that $k_{\text{w}}/k$ can reach $5.3$ which warrants a fast gate. Meanwhile, the $\pi-4\pi-\pi$ pulse sequence for the target qubit in Fig.~\ref{figure03} occurs during the wait time, $t\in[t_\pi,t_\pi+t_{\text{w}})$, i.e., within a duration of $t_{\text{w}}=t_{\text{gate}}k / (k_{\text{w}}-k)$. The fact $t_{\text{w}}/t_{\text{gate}}=k / (k_{\text{w}}-k)<1$ means that the Rabi frequency $\Omega_{1}'$ used in the target qubit should be larger than $\Omega_{1}$ by a factor of $ (k_{\text{w}}-k)/k$. As an example, we consider
\begin{eqnarray}
  \Omega_{1}'&=& \Omega_{1}(k_{\text{w}}-k)/k,\nonumber\\
  \Omega_2' &=& 2\Omega_{1}'(k_{\text{w}}-2k)/k ,\label{Omega12prime}
\end{eqnarray}
which indicates that although a large $k_{\text{w}}/k$ warrants a small $t_{\text{gate}}$, it requires larger $\Omega_{1(2)}'$ to work. Unfortunately, the dipole moment between $|e_{1(2)}\rangle$ and $|r_{1(2)}\rangle$ is small, so the two-photon Rabi frequencies $\Omega_{1(2)}$ and $\Omega_{1(2)}'$ can not be large. With rare exceptions~\cite{Gaetan2009}, they were smaller than $2\times2\pi$~MHz in typical experiments~\cite{Wilk2010,Isenhower2010,Zhang2010,Maller2015,Zeng2017,Levine2018,Picken2018}. Although a recent experiment has used Rabi frequencies up to $4.6\times2\pi$~MHz~\cite{Graham2019}~(or, $3.5\times2\pi$~MHz~\cite{Levine2019}) for exciting the $66S_{1/2}$ state of $^{133}$Cs~(or, $70S_{1/2}$ of $^{87}$Rb), we will be conservative and consider Rabi frequencies smaller than $2\times2\pi$~MHz. With this compromise, we identify from the data in Refs.~\onlinecite{Sansonetti2006,Sansonetti2009}, a set of parameters, namely, $5P_{3/2}$ as $|e_1\rangle$ and $6P_{1/2}$ as $|e_2\rangle$ for $^{87}$Rb, where $k_{\text{w}}/k=2.4767$. With $\Omega_1/2\pi=1.35$~MHz, Eq.~(\ref{Omega12prime}) results in a set of Rabi frequencies all below $2\times2\pi$~MHz, where $t_{\text{gate}} =2.3~\mu$s. For qubits in a cryogenic~(room) temperature of $4$~(300)~K, the lifetime of an $s-$orbital Rydberg states $|r_{1(2)}\rangle$ is around 1.2~(0.3)~ms~\cite{Beterov2009}, leading to a decay error on the order of $10^{-4}~(10^{-3})$ for the method of estimate in Refs.~\onlinecite{Saffman2005,Zhang2012}. 

The remaining gate error comes from the blockade error and a residual Doppler dephasing, as our scheme is perfect only for the impossible case of infinitesimal $kv/\Omega_{1(2)}$ and $kv/\Omega_{1(2)}'$. The input states $|01\rangle$ and $|10\rangle$ are studied according to the pumping scheme in Figs.~\ref{figure03}(a) and~\ref{figure03}(b), respectively. The input state $|11\rangle$, which involves Rydberg interaction, is analyzed below. During $t\in[0,~t_\pi)$, the transition $|11\rangle\leftrightarrow|r_11\rangle$ occurs with Rabi frequency $\Omega_1 e^{ik(z_{0,c}+v_ct)}$, where $z_{0,c(t)}$ is the initial $\mathbf{z}$-coordinate and $v_{c(t)}$ is the velocity of the control~(target) qubit. The atoms are mainly in the state $|r_11\rangle$ but can also have some population in $|11\rangle$ at $t=t_\pi$. During $t\in[t_\pi,t_\pi+t_{\text{w}})$, the $|11\rangle$-component evolves according to the single-particle picture of Fig.~\ref{figure03}(a), while $|r_11\rangle$ is coupled with other states according to the following Hamiltonian
 \begin{eqnarray}
  \centering
\frac{1}{2}  \left(
  \begin{array}{cccccc}
    2V_{22}& \kappa_2^{\ast} & \kappa_2'^{\ast}& 0&  0& 0 \\
  \kappa_2& 2V_{12}& 0 &  \kappa_2'^{\ast} & 0&  0 \\
   \kappa_2'&0 &2V_{12}& \kappa_2^{\ast}& \kappa_1' & 0 \\
   0 & \kappa_2'&  \kappa_2&2V_{11}& 0& \kappa_1'\\
   0& 0 & \kappa_1'^{\ast} & 0&0&  \kappa_2^{\ast} \\
  0&  0& 0 & \kappa_1'^{\ast}& \kappa_2&0\\    
    \end{array}
  \right),
  \label{hamiltonian1r}
 \end{eqnarray}
 which is written in the basis $\{|r_2r_2\rangle$$,~|r_1r_2\rangle$$,~|r_2r_1\rangle$,~$|r_1r_1\rangle$$,~|r_21\rangle$,$~|r_11\rangle\}$, where $\kappa_2\equiv \Omega_{\text{2}} e^{ik_{\text{w}}(z_{0,c}+v_ct)}$, $\kappa_1'\equiv\Omega_{\text{1}}' e^{ik_{\text{}}(z_{0,t}+v_tt)}$, and $\kappa_2'\equiv\Omega_{\text{2}}' e^{ik_{\text{w}}(z_{0,t}+v_tt)}$. The van der Waals interactions~\cite{Walker2008} are given by $[V_{11},~V_{12},~V_{22}]= [C_6(r_1r_1),~C_6(r_1r_2),~C_6(r_2r_2)]/L^6$, where $L$ is the two-qubit spacing. In Eq.~(\ref{hamiltonian1r}), $\Omega_{\text{2}}$ is nonzero during the whole wait time, while $\Omega_{\text{1}}'$ is nonzero during the two $\pi$ pulses in Fig.~\ref{figure03}(a), and $\Omega_{\text{2}}'$ is nonzero only during the $4\pi$ pulse for the target. After the wait time, the final $\pi$ pulse of Fig.~\ref{figure03}(b) is used in the control qubit, completing the gate sequence.

 \begin{table}
  \centering
  \begin{tabular}{|c|c|c|c|}
    \hline
  $T~(\mu$K)    & 10 & 100& 200 \\\hline
 $\overline{E_{\text{ro}}}$~(This work) & $4.74\times10^{-4}$ &  $1.07\times10^{-2}$&  $3.11\times10^{-2}$  \\\hline 
 $\overline{E_{\text{ro}}}$~(Traditional)  & $1.87\times10^{-2}$ &  $1.52\times10^{-1}$&  $2.48\times10^{-1}$  \\\hline    
  \end{tabular}
  \caption{ Rotation error of a blockade gate with $(\Omega_{1},\Omega_{2}, \Omega_{1}', \Omega_{2}')/2\pi=(1.35,1.29,1.99,1.90)$~MHz in our method, and with the traditional method where all Rabi frequencies are $1.35\times2\pi$~MHz. Both gates have the same duration and the same wait time of $t_{\text{w}}=1.55~\mu$s.   \label{table2}  }
  \end{table}


 To show that our method is suitable for suppressing the motional dephasing, we choose a condition where the blockade error is negligible so that the rotation error is dominated by the residual Doppler dephasing, which is ensured 
by large blockade interactions. For illustration, we choose an $s-$orbital Rydberg state of $^{87}$Rb with a principal value $99~(100)$ as $|r_{1(2)}\rangle$, where the $C_6$ coefficients are given by $[C_6(r_1r_1),~C_6(r_1r_2),~C_6(r_2r_2)]/2\pi=[50,68,56]$THz~$\mu m^6$. For $L=8~\mu$m, we ignore the change in $V_{11},~V_{12}$, and $V_{22}$ during the gate sequence arising from the atomic drift because the blockade error is insensitive to their change.
As shown in Appendix~\ref{appA}, we define the rotation error as in Ref.~\onlinecite{Pedersen2007}, and calculate the expected rotation error, $\overline{E_{\text{ro}}}$, by averaging over $10^4$ sets of atomic speeds $(v_{c},~v_t)$, where $v_{c(t)}$ takes $100$ values equally distributed from $-0.5$ to $0.5$~m/s because the atomic speed has little chance to exceed $0.5$~m/s for $T<0.2$~mK. The simulated $\overline{E_{\text{ro}}}$ is shown in Table~\ref{table2} with $T=10,~100$, and $200~\mu$K. For an unbiased comparison, Table~\ref{table2} also shows $\overline{E_{\text{ro}}}$ for a traditional gate with all Rabi frequencies equal to $\Omega_1$, the same wait time $t_{\text{w}}$~(i.e., the $2\pi$ pulse for the target is shorter than $t_{\text{w}}$), and a blockade interaction equal to $V_{22}$. It shows that $\overline{E_{\text{ro}}}$ in our gate is smaller than that in the traditional gate by $39,14$, and $8$ times when $T$ is $10,100$, and $200~\mu$K, respectively. The rapid increase of $\overline{E_{\text{ro}}}$ with $T$ in our method arises from that $k_{\text{w}}v_{\text{rms}}/\Omega_2$ is $0.048,0.15,$ and $0.21$ for $T=10,100$, and $200~\mu$K, respectively, which confirms the condition of $kv\ll\Omega$ for our theory as studied in Fig.~\ref{figure02}. Because the four Rabi frequencies used in Fig.~\ref{figure03} 
cannot be equal, if we restrict all of them to be below $2\times2\pi$~MHz, some of them will be quite small so that the condition $kv\ll\Omega$ can not be well satisfied. However, the qubits were usually cooled to around or below $10~\mu$K in most recent experiments~\cite{Levine2018,Picken2018,Graham2019}, thus the theory in this work is sufficient for high-fidelity quantum control.

\section{Conclusions}\label{sec04}
In this work, we show a $\pi-2\mathcal{N}\pi-\pi$ protocol for removing the motional dephasing in the ground-Rydberg transition of a neutral atom in free flight. It can be used, in principle, for excitation of any $s$ or $d-$orbital Rydberg state via a two-photon transition. The protocol relies on a linear phase modulation of the flying atom in response to a transition between two nearby Rydberg states. It works well when $kv$ is much smaller than the magnitude of the Rabi frequency, where $k$ is the wavevector of the laser fields and $v$ is a typical velocity of the atom. This condition is easily satisfied for atoms cooled to around $10~\mu$K, where a high-fidelity Rydberg blockade gate is realizable because of negligible motional dephasing. This simple method for suppressing the dephasing paves the way to neutral-atom quantum computing with minimal technical resources and is applicable for many quantum control tasks with neutral Rydberg atoms.

\section*{ACKNOWLEDGMENTS}
The author thanks Yan Lu for insightful discussions. This work was supported by the National Natural Science Foundation of China under Grant No. 11805146.

\appendix{}
\section{Gate error}\label{appA}
Ideally, the Rydberg blockade gate maps the input states according to $\{|00\rangle,~|01\rangle,~|10\rangle,~|11\rangle\}\rightarrow \{|00\rangle,~-|01\rangle,~-|10\rangle,~-|11\rangle\}$~\cite{PhysRevLett.85.2208}. Due to the finiteness of $\Omega_{1(2)}~(\Omega_{1(2)}')$ and the Rydberg-state decay, the actual gate in the matrix form becomes,
\begin{eqnarray}
 \mathscr{U} &=& \left(
  \begin{array}{cccc}
    1& 0 & 0&0\\
    0 & a &0&0\\
    0 &0 & b&0\\
    0& 0 & 0&c\\   
    \end{array} 
  \right) ,\label{realgate01}
  \end{eqnarray}
where $a$ and $b$ can deviate from $-1$ due to Doppler dephasing, $c$ can acquire errors from both Doppler dephasing and the finiteness of the blockade interaction, and the magnitudes of $a,~b$, and $c$ become smaller than 1 due to Rydberg-state decay. The fidelity error is then given by
\begin{eqnarray}
 E &=& E_{\text{ro}} + E_{\text{decay}}.
\end{eqnarray}
Here the rotation error is~\cite{Pedersen2007}
\begin{eqnarray}
 E_{\text{ro}} &=& 1-\frac{1}{20}\left[  |\text{Tr}(U^\dag \mathscr{U})|^2 + \text{Tr}(U^\dag \mathscr{U}\mathscr{U}^\dag U ) \right], \label{fidelityError01}
\end{eqnarray}
where $\mathscr{U}$ is evaluated by using the Hamiltonian dynamics with the Rydberg-state decay ignored, $U=$ diag$\{1,-1,-1,-1\}$, and the Rydberg-state decay can be approximated by~\cite{Zhang2012} 
\begin{eqnarray}
E_{\text{decay}}&=&[T_{\text{r}}(01)+T_{\text{r}}(10) +T_{\text{r}}(11)]/(4\tau),\label{fidelityError02}
\end{eqnarray}
where $T_{\text{r}}(ab)$ is the time for the input state $|ab\rangle$ to be in a single-Rydberg state; the time for the input state $|11\rangle$ to be in a two-atom Rydberg state $|r_\alpha r_\beta\rangle$ throughout the gate sequence is tiny and neglected. If we let $\tau$ be the smaller of the lifetimes of the Rydberg states $|r_1\rangle$ and $|r_2\rangle$, the Rydberg-state decay will be overestimated slightly. For the parameters in the main text, our estimate shows that the decay error is on the order of $10^{-4}~(10^{-3})$ for qubits in a cryogenic~(room) temperature of $4$~(300)~K.

The evaluation of the average rotation error, $\overline{E_{\text{ro}}}$, involves a two-dimensional integration over the distribution of $v_c$ and $v_t$, both of which are sampled from the Maxwell distribution $\mathscr{G}(v)$. $\mathscr{G}(v)$ is Gaussian in one dimension,
\begin{eqnarray}
  \mathscr{G}(v) &=& \sqrt{\frac{m}{2\pi k_BT}}\text{exp}\left(-\frac{mv^2}{2k_BT}\right),
\end{eqnarray}
where $k_B$ is the Boltzmann constant, $T$ is the atomic temperature, and $m$ is the mass of the qubit. Because of the constant drift of the two qubits, very small time step is required in the simulation for each $v$, and thus an accurate convergence of this integration requires long time of simulation. We approximate the two-dimensional integration by
\begin{eqnarray}
\overline{E_{\text{ro}}}\approx \frac{\sum_{v_c}\sum_{v_t} E_{\text{ro}}(v_c,~v_t)\mathscr{G}(v_c) \mathscr{G}(v_c) }{ \sum_{v_c}\sum_{v_t} \mathscr{G}(v_c) \mathscr{G}(v_c)},\label{roterror02}
\end{eqnarray}
where the sum is over $10^4$ sets of speeds $(v_{c},~v_t)$, where $v_{c(t)}$ applies $100$ values equally distributed from $-0.5$ to $0.5$~m/s because the atomic speed has little chance to be over $0.5$~m/s for $T<0.2$~mK. We have compared results between the calculation by Eq.~(\ref{roterror02}) and a rigorous integration for shorter pumping time and found negligible relative error. By using Eq.~(\ref{roterror02}), the third row of Table~\ref{table1} shows three rotation errors of the traditional gate, where the data with $T=100$ and $200~\mu$K matches well with the corresponding data of Fig.~3 of Ref.~\cite{Saffman2011}, in which a similar wavevector of $2\pi(1/480-1/780)$nm$^{-1}$ was used. For $T=10~\mu$K, the rotation error of the traditional gate in Table~\ref{table1} is about 2 times larger than that in Ref.~\cite{Saffman2011}. A likely reason is that, for colder qubits, the dephasing during gap time is comparable to that during the pumping time, while the estimate in Ref.~\cite{Saffman2011} only accounts for dephasing during the gap time.

%


\end{document}